\documentclass[a4paper,11pt]{article}

\usepackage{lineno}
\usepackage{amssymb}
\usepackage{amsfonts,amsbsy,graphics,epsfig}
\usepackage{graphicx}
\usepackage{amsmath}
\usepackage{color}
\usepackage{multirow}
\usepackage{arydshln}

\usepackage{booktabs}
\usepackage{url}

\usepackage[dvipsnames]{xcolor}
\usepackage{wasysym}
\usepackage{natbib}
\usepackage{cleveref}
\usepackage{adjustbox}
\modulolinenumbers[1]

\usepackage{authblk}

\title{Spatial Autoregressive Models for Scan Statistic}

\author[1]{Mohamed-Salem AHMED}
\affil[1]{Univ. Lille, CHU Lille, EA 2694 – Evaluation des technologies de sant\'e et des pratiques m\'edicales, F-59000 Lille, France.}
\author[2]{Lionel CUCALA}
\affil[2]{IMAG, Universit\'e de Montpellier, CNRS, Montpellier, France.}
\author[1]{Micha\"el GENIN}

\date{ }

\begin{document}
  \maketitle
  \begin{abstract}
Spatial scan statistics are well-known methods for cluster detection and are widely used in epidemiology and medical studies for detecting and evaluating the statistical significance of disease hotspots. For the sake of simplicity, the classical spatial scan statistic assumes that the observations of the outcome variable in different locations are independent, while in practice the data may exhibit a spatial correlation.
In this article, we use spatial autoregressive (SAR) models to account the spatial correlation in parametric/non-parametric scan statistic. Firstly, the correlation parameter is estimated in the SAR model to transform the outcome into a new independent outcome over all locations. Secondly, we propose an adapted spatial scan statistic based on this independent outcome for cluster detection. 
A simulation study highlights the better performance of the proposed methods than the classical one in presence of spatial correlation in the data. The latter shows a sharp increase in Type I error and false-positive rate but also decreases the true-positive rate when spatial correlation increases. Besides, our methods retain the Type I error and have stable true and false positive rates with respect to the spatial correlation. The proposed methods are illustrated using a spatial economic dataset of the median income in Paris city. In this application, we show that taking spatial correlation into account leads to the identification of more concentrated clusters than those identified by the classical spatial scan statistic.\\

\end{abstract}

\section{Introduction}
\label{}

In many fields of science, cluster detection methods are useful tools for objectively identifying aggregations of events in time and/or space and for determining their statistical significance. 

Over the last few decades, several cluster detection methods have been developed. In particular, spatial scan statistics  \citep[originally proposed by ][for count spatial data]{kulldorff1997spatial} are powerful methods for detecting spatial clusters with a variable scanning window size and in the absence of pre-selection bias, and then testing the clusters' statistical significance. Following on from Kulldorff's initial work, several researchers have adapted spatial scan statistics to continuous spatial data. Many of them use a parametric approach and specify the distribution of the data: exponential \citep[][]{huang2007spatial},  normal \citep[][]{kulldorff2009scan, huang2009weighted}, Weibull  \citep[][]{bhatt2014spatial}, ... Others use a nonparametric approach based on moments \citep[][]{cucala2014distribution} or ranks \citep[][]{jung2015nonparametric}.

A common assumption in the literature of spatial scan statistics is that spatial data is composed of independent observations, for reasons of model simplicity. However, spatial data are usually characterized by the notion of spatial correlation, which makes the previous hypothesis too simplistic and inadequate. Particularly, it is reasonable to expect some positive correlation between nearby locations in  epidemiological or environmental studies. For instance, it is not surprising to say that the air quality in a given location depends on those measured in neighboring sites. Additionally, one can imagine that there is some pollution source  which is behind the diffusion. Hence, the fact that spatial sites near this source are more polluted than those further away could be due solely to the effect of spatial correlation. Thus, in such situations, it is important that spatial scan statistics detect only the source of pollution and not a larger spatial cluster characterized by a strong positive spatial correlation.

In the literature, a small number of studies \citep[][]{loh2007accounting,lin2014generalized,leespatial} have focused on taking spatial correlation into account within spatial scan statistical methods and studying their behaviour in this situation. For instance, \cite{loh2007accounting} showed in theoretical as well as practical point of view that ignoring the spatial correlation leads to an increased rate of false positive. Briefly, these works proposed  modified spatial scan statistics allowing to integrate residual spatial correlation present in the data. Residual spatial correlation is principally due to spatial
heterogeneity or  omission of some spatial correlated latent variables that are related to the study but not included in the data (omission of some spatial correlated confounding factor). The reader may refer to Chapter 2 in \cite[][]{lesage2009introduction} for more discussion about the different kinds of spatial dependence. However, these  previous studies do not allow the spatial correlation to be taken into account when studying contagious phenomena, such as infectious diseases, which are usually characterized by a spatial correlation affecting the dependent variable of interest.
\\ 

In this paper, we focused on continuous spatial data in which the spatial correlation is integrated into the dependent variable using spatial autoregressive models (SAR). SAR models were introduced by \cite{Cliff} and are usually used in the literature of spatial econometrics \cite[see Section 2.1 in ][]{lesage2009introduction}. By combining \cite{jung2009generalized}'s approach, expressing spatial scan statistics in terms of generalized linear models, with SAR models, we integrate the spatial correlation in the proposed SAR scan statistics through some a known spatial weights matrix and  an unknown scalar named the spatial autoregressive parameter. The latter allows to control the intensity of the spatial correlation while the spatial weights matrix allows to describe the spatial interactions between locations. We showed that the SAR scan statistic is equivalent to using a conventional spatial scan statistic after adjusting the initial dependent variable for spatial correlation.\\
Firstly, we developed some estimation procedure to build the spatial autoregressive parameter based on a quasi-maximum likelihood (QML) method proposed by \cite{lee2004asymptotic}. Secondly, we used this QML estimator to construct some transformation of the initial dependent variable removing the effect of spatial correlation. Finally, as the new transformed dependent variable satisfies the independence assumption of classical spatial scan statistics, we suggest to use the Gaussian-based spatial scan statistic proposed by \cite{kulldorff2009scan} in case of normally distributed initial data and the distribution-free scan statistic proposed by \cite{cucala2014distribution} otherwise. \\

The present article is organized as follows. Section \ref{Methodology} describes the methodology of the classical spatial scan statistic, the two proposed SAR models and  presents an estimation procedure of the spatial autoregressive parameter. Section \ref{Simulation_study} presents both the design and the results of a simulation study. In Section \ref{Real_data} we apply the SAR scan statistics to economic data and the detection of clusters of high and low income in the city of Paris. Lastly, the results are discussed in Section \ref{Discussion}.

\section{Methodology}\label{Methodology}

Let consider that at each location $s_i$ (one of $n$ different spatial locations $s_1,\ldots,s_n$ included in $D\subset \mathbb{R}^2$), we observe a continuous outcome variable $Y_i$ ($1\le i\le n$). A spatial scan statistic usually denotes the maximum concentration observed among a collection of potential clusters denoted by $\mathcal{C}=\{C_k \subset D,\, k=1,2,... \}$. It is used as a test statistic for areas in which the concentration might be abnormally high or abnormally low \citep[][]{Cressie1977}. Without loss of generality and in line with Kulldorff's work \citep[][]{kulldorff1997spatial}, we shall focus on variable-size circular clusters. Hence, the set of potential clusters $\mathcal{C}$ is built so that (i) each potential cluster is centered at a particular location, and (ii) the radius is limited so that the corresponding cluster cannot cover more than 50\% of the studied region. It should be noted that many other configurations such as elliptical clusters \citep[][]{kulldorff2006elliptic} and graph-based clusters \citep[][]{cucala2013spatial} have been suggested. The scan statistic for continuous spatial data is detailed in the following subsection. 

\subsection{The Gaussian scan statistic for continuous data}

 \cite{kulldorff2009scan} introduced a Gaussian-based scan statistic to detect clusters when dealing with univariate continuous data. It relies on the likelihood ratio between two hypotheses: the $Y_i$'s are supposed to be normally-distributed and independent but the null hypothesis considers equal means and variances whereas the alternative hypothesis considers equal variances but different means inside and outside the potential cluster. Following the approach proposed by  \cite{jung2009generalized} and for a given potential cluster $C_k \in \mathcal{C}$, the $Y_i$'s can be expressed in terms of linear model as follows:
\begin{equation}
Y_i =\alpha + \delta_k \xi_i^{(k)}+ \epsilon_i, \qquad i=1,\ldots,n
\label{modeliid}
\end{equation}
where $\xi_{i}^{(k)}$ is a binary covariate equal to $1$ if  the location $s_i$  belongs to $C_k$, $0$ otherwise. $\alpha$ is the intercept, $\delta_k$ refers to the intensity of the cluster, and the disturbances $\epsilon_i,\, i=1,\ldots,n,$ are assumed to be independent normally distributed with mean zero and unknown variance $\sigma^2$. Based on this model, the test hypotheses can be expressed as follows:
\begin{equation*}
\left\{
\begin{array}{lcl}
\mathcal{H}_0 & : &   \delta_{k}=0 \\ 
& & \\ 
\mathcal{H}_1 &: & \delta_{k}\neq 0
\end{array}
\right.
\end{equation*}
Under $\mathcal{H}_0$, it means that $Y_i \sim \mathcal{N}(\alpha,\sigma) $ for all locations $s_i$ while, under $\mathcal{H}_1$,  $Y_i \sim \mathcal{N}(\alpha+\delta_k,\sigma)$ for $s_i$ inside $C_k$ and $Y_i \sim \mathcal{N}(\alpha,\sigma)$ for $s_i$ outside $C_k$.

The log-likelihood ratio (LLR) related to these hypotheses is defined by: 
\begin{eqnarray}
\mathrm{LLR}(Y;C_k)&=&L_n\left(Y;\widehat{\alpha}_{k},\widehat{\sigma^2}_{k},\widehat{\delta}_k\right)-L_n\left(Y;\widehat{\alpha},\widehat{\sigma^2}, 0\right)
=\frac{n}{2}\left(\log\left(\widehat{\sigma^2}\right)-\log\left(\widehat{\sigma^2}_{k}\right)\right),
\label{LLRiid}
\end{eqnarray} 
where $L_n$ refers to the log-likelihood function of the model (\ref{modeliid}): 
\begin{equation}
L_n\left(Y;\alpha,\sigma^2, \delta_k\right)=-\frac{n}{2}\log(\sigma^2) -\frac{1}{2\sigma^2}\sum_{i=1}^{n}\left( {Y}_i-\alpha-\delta_{k}{\xi}_i^{(k)}\right)^{2},
\label{LLiid}
\end{equation}
and $\widehat{\alpha}_{k}$, $\widehat{\sigma^2}_{k}$, and $\widehat{\delta}_k$ denote respectively the maximum likelihood estimators (MLE) of $\alpha$, $\sigma^2$, and  $\delta_{k}$ under $\mathcal{H}_1$ while $ \widehat{\alpha}$ and $\widehat{\sigma^2}$ refer to the MLEs of $\alpha$ and $\sigma^2$ under $\mathcal{H}_0$. It should be noted that the LLR defined in (\ref{LLRiid})  is equal to the LLR described by \cite{kulldorff2009scan}, see for instance the expression of the MLE $\widehat{\sigma^2}_{k}$ in the Appendix.

The Most likely cluster (MLC) is then defined as the potential cluster $C_k$ that maximizes the LLR:
\begin{equation}
\widehat{C}=\mathrm{argmax}_{C_k \in \mathcal{C}}\{\mathrm{LLR}(Y;C_k)\}.
\end{equation}
Hence, the Gaussian spatial scan statistic is defined as the LLR associated with the MLC:
\begin{equation}
\lambda_G = \max_{C_k \in \mathcal{C}}\{\mathrm{LLR}(Y;C_k)\}.
\end{equation}

\subsection{Parametric SAR scan statistic}
The classical spatial scan statistics assume that the $Y_i$'s are spatially independent while, in some cases, this assumption may be violated, especially when the spatial data is generated through contagious phenomenon. The latter is characterized by the fact that the intensity of the outcome variable will depend on the distance (either euclidian or based on any neighbouring network) from the source that is behind the diffusion. In this situation, it is more appropriate to integrate the spatial correlation within the outcome variable rather than in the residuals. Therefore, this can be achieved by using the following spatial version of the previous model (\ref{modeliid}): 
\begin{equation}
Y_i =\sum_{j=1, \,j\neq i}^{n}f(d_{ij},\theta)Y_j+\alpha + \delta_k \xi_i^{(k)}+ \epsilon_i
\label{modele1}
\end{equation}
where  $f(\cdot, \cdot)$ is a function of distance $d_{ij}$ between locations $s_i$ and $s_j$, parametrized by a vector of  coefficients $\theta$. This model implies that $Y_i$'s are composed of i) a neighboring effect ii) a baseline effect, and  iii) a clustering effect.\\
For sake of simplicity, one can replace the function of distance  $f(\cdot , \cdot)$ by some spatial  weights matrix $W_n$ whose elements are such that $w_{ii}=0$ and the $w_{ij}$'s are usually considered as inversely proportional to $d_{ij}$. In addition, for interpretative reason, the spatial weights matrix is often row-standardized in order to have elements with row sum equal to one \citep{anselin2013spatial}. This allows to have spatial weights matrix with elements between $0$ and $1$ and facilitates the interpretation of the first term in the right hand of (\ref{modele1}) as an averaging of neighboring values. Such consideration leads to define the model $(\ref{modele1})$ as a particular version of the well-known spatial autoregressive (SAR) model \citep{Cliff}:
\begin{equation}
Y_i =\rho^{*}\sum_{j=1}^{n}w_{ij}Y_j+\alpha + \delta_k \xi_i^{(k)}+ \epsilon_i
\label{sar}
\end{equation}
where $\rho^{*}$ is the spatial autoregressive parameter explaining the intensity of correlation between outcome observations. \\
Let $\mathbf{1}$ be the $n \times 1$ ones vector, $\mathbf{Y}$, $\xi^{(k)}$ and $\mathbf{\epsilon}$ are the $n\times 1$ vectors with elements $Y_i$, $\xi_i^{(k)}$ and $\epsilon_i$ respectively. One can rewrite the SAR model (\ref{sar}) as follows:   
\begin{equation}
(I_n-\rho^{*} W_n)\mathbf{Y}=\alpha\mathbf{1}+\delta_k\mathbf{\xi}^{(k)} +\mathbf{\epsilon},
\label{sarTrans}
\end{equation}
where $I_n$ is the identity matrix  and $\mathbf{Y}^{(\rho^{*})}=(I_n-\rho^{*} W_n)\mathbf{Y}$ denotes the spatially filtered version of $\mathbf{Y}$. This latter denotes the transformation of the initial outcome variable removing the effect of spatial correlation. It should be noted that the $Y^{(\rho^{*})}_i$'s,  $i=1,\ldots,n$, are independent and, under $\mathcal{H}_0$, $Y^{(\rho^{*})}_i \sim \mathcal{N}(\alpha,\sigma) $ for all locations $s_i$ while, under $\mathcal{H}_1$,  $Y^{(\rho^{*})}_i \sim \mathcal{N}(\alpha+\delta_k,\sigma)$ for $s_i$ inside $C_k$ and $Y^{(\rho^{*})}_i \sim \mathcal{N}(\alpha,\sigma)$ for $s_i$ outside $C_k$.
Therefore, if this vector was available, the Gaussian spatial scan statistic of \cite{kulldorff2009scan} should be applied to $\mathbf{Y}^{(\rho^{*})}$ instead of $\mathbf{Y}$ which is affected by the spatial correlation. Unfortunately, since the real spatial autoregressive parameter $\rho^*$ is unknown, $\mathbf{Y}^{(\rho^{*})}$  has to be approximated by replacing the spatial autoregressive parameter by some consistent estimator $\widehat{\rho}$ that we will describe later on. Let $\mathbf{Y}^{(\widehat{\rho})}=(I_n-\widehat{\rho} W_n)\mathbf{Y} $. We can now define the parametric SAR scan statistic:
\begin{equation}
\lambda_{P-SAR} = \max_{C_k \in \mathcal{C}}\{\mathrm{LLR}(\mathbf{Y}^{(\widehat{\rho})};C_k)\}.
\end{equation}

\subsection{Non-parametric SAR scan statistic}

An alternative to the Gaussian-based spatial scan statistic of \cite{kulldorff2009scan}, named as the distribution-free scan statistic, has been proposed by \cite{cucala2014distribution}. Contrary to classical variable window scan methods, the concentration index to maximize on the set of potential clusters $\mathcal{C}$ is not based on a likelihood ratio and thus does not depend on any specific probability distribution. 

Let consider the spatially filtered version of $\mathbf{Y}$, $\mathbf{Y}^{(\rho^{*})}$, which, as already said in Subsection 2.2, satisfies the i.i.d. assumption under $\mathcal{H}_0$. The distribution-free concentration index applied to $\mathbf{Y}^{(\rho^{*})}$ is 
$$ I^{(\rho^{*})}(C_k)= \frac{\sqrt{n_k (n-n_k) }}{\sqrt{n}} \left| \frac{1}{n_k} \sum_{\{i:s_i \in C_k \}}\mathbf{Y}_i^{(\rho^{*})} - \frac{1}{n-n_k} \sum_{\{i:s_i \notin C_k \}}\mathbf{Y}_i^{(\rho^{*})}\right| $$
where $n_k=\sum_{i=1}^n \xi_{i}^{(k)}$ is the number of locations in $C_k$. This concentration index measures the difference of values observed inside $C_k$ and outside $C_k$ and, under $H_0$, it has null expectation and its variance does not depend on $n_k$. It has been shown that it is very powerful to detect clusters whatever the underlying distribution.

Since the real spatial autoregressive parameter $\rho^*$ is unknown, we might again replace it by $\widehat{\rho}$. The distribution-free concentration index applied to $\mathbf{Y}^{(\widehat{\rho})}$ is 
$$ I^{(\widehat{\rho})}(C_k)= \frac{\sqrt{n_k (n-n_k) }}{\sqrt{n}} \left| \frac{1}{n_k} \sum_{\{i:s_i \in C_k \}}\mathbf{Y}_i^{(\widehat{\rho})} - \frac{1}{n-n_k} \sum_{\{i:s_i \notin C_k \}}\mathbf{Y}_i^{(\widehat{\rho})} \right| .$$
Hence, we also propose a nonparametric SAR scan statistic defined as:
\begin{equation}
\lambda_{NP-SAR} = \max_{C_k \in \mathcal{C}}\{\mathrm{I^{(\widehat{\rho})}}(C_k)\}.
\end{equation}

\subsection{Estimation of the spatial autoregressive parameter}
It should be noted that i) the spatial correlation assumption  is considered under both hypotheses $\mathcal{H}_0$ and $\mathcal{H}_1$, and  ii) the intensity of spatial correlation ($\rho^*$) should not vary between these two hypotheses because it depends on the spatial structure of the studied region rather than the  clustering hypotheses. However, there is only  one autoregressive parameter $\rho^*$ and this should be estimated under the true hypothesis among $\mathcal{H}_0$ and all alternatives hypotheses related to $C_k\in \mathcal{C}$, i.e. the hypothesis under which the observations were generated by model (\ref{sar}). Therefore, for each candidate hypothesis ($\mathcal{H}_0$ or $\mathcal{H}_1$), we need to assess the ability of model (\ref{sar}) to describe  the relationship between observations. Intuitively, one needs to compare  the "best" SAR model (\ref{sar}) (maximizing the log-likelihood over all potential clusters $ C_k \in \mathcal{C}$) with the model (\ref{sar}) built under $\mathcal{H}_0$. This model selection between these two models is performed using the Bayesian information criteria (BIC). \\
\noindent  Firstly, for each $C_k \in \mathcal{C}$, the SAR model (\ref{sar}) has the following log-likelihood: 
\begin{equation}
	L_n(\mu,\sigma^2, \delta_k, \rho)=\log\left(\det\left(I_n-\rho W_n\right)\right)-\frac{n}{2}\log(\sigma^2) -\frac{1}{2\sigma^2}\sum_{i=1}^{n}\left( {Y}^{(\rho)}_i-\alpha-\delta_{k}{\xi}_i^{(k)}\right)^{2}. 
	\label{LL.H1}
\end{equation}
Let  $\widehat{\alpha}_{k}$, $\widehat{\sigma^2}_{k}$, $\widehat{\delta}_k$, and  $\widehat{\rho}_k$ be the MLEs under $\mathcal{H}_1$ of  $\alpha$, $\sigma^2$, $\delta_{k}$, and  $\rho^{*}$ respectively (see \cite{lee2004asymptotic} for more details). Therefore, the BIC associated to this candidate model is:
\begin{equation*}
\mathrm{BIC}_{k}=p\log(n)-2L_n\left(\widehat{\alpha}_{k},\widehat{\sigma^2}_{k},\widehat{\delta}_k, \widehat{\rho}_k\right),
\end{equation*}
where $p$ corresponds to the number of parameters to estimate (here, $p=4$).\\
Hence,  the "best" model (\ref{sar}) over all potential clusters  $\mathcal{C}$ is the one that minimizes the $\mathrm{BIC}_{k}$'s:
\begin{equation*}
\mathrm{BIC}_{k^{*}}=\displaystyle\min_{C_k\in \mathcal{C}}\mathrm{BIC}_{k}
\end{equation*}
and let denote by $\widehat{\rho}_{k^{*}}$ the MLE  of  $\rho^{*}$ associated to this model. \\
\noindent Secondly, the BIC associated to model (\ref{sar}) under the null hypothesis is: 
\begin{equation*}
\mathrm{BIC}_{0}=p_0\log(n)-2L_n\left(\widehat{\alpha}_0,\widehat{\sigma^2}_0,0, \widehat{\rho}_0\right),
\end{equation*}
where, $p_0=3$, $\widehat{\alpha}_0$, $\widehat{\sigma^2}_0$, and $\widehat{\rho}_0$ are the MLEs under $\mathcal{H}_0$ of $\alpha$, $\sigma^2$, and  $\rho^{*}$ respectively. \\ 
Finally, in order to assess the relative merits between model $(\ref{sar})$ under $\mathcal{H}_0$ and the "best" model (\ref{sar}) under $\mathcal{H}_1$, we follow the rules of thumb proposed  by \cite{raftery1995bayesian}. Briefly, these rules of thumb rate the BIC's difference $\Delta= \mathrm{BIC}_{0}-\mathrm{BIC}_{k^{*}}$ as "positive", "strong", or "very strong" whether it lies in one of these levels $]2, 6]$, $]6, 10]$, or $]10,+ \infty[$, respectively. According to these rough guidelines, we define the MLE of $\rho^{*}$ by:
\begin{equation*}
\widehat{\rho}=\left\{
\begin{array}{l l}
\widehat{\rho}_{k^{*}} & \mbox{if} \quad  \Delta>10,\\ 
\widehat{\rho}_{0} & \mbox{otherwise}
\end{array}
\right.
\end{equation*}
It should be noted that $\Delta>10$ corresponds to the case of essentially  no support between the two models \cite[see][]{burnham2004multimodel}.

\subsection{Computing the significance}

In the following, let $\lambda$ refer to the scan statistic of one of the three previous methods ($\lambda_G$, $\lambda_{P-SAR}$ or $\lambda_{NP-SAR}$). Since the distribution of $\lambda$ under $\mathcal{H}_0$ does not have a closed form, \cite{kulldorff2009scan} suggested to evaluate the statistical significance of the MLC by using Monte-Carlo simulations. Let generate $M$ randomly permuted data sets (the initial dataset $Y$ when $\lambda=\lambda_G$ or the spatially filtered dataset $\mathbf{Y}^{(\widehat{\rho})}$ otherwise) and $\lambda^{(1)}, \dots ,\lambda^{(M)}$ be the observations of the scan statistic on these data sets. Then the p-value of the $\lambda$ observed in the real data is defined by $1-R/(M+1)$, where $R$ is the rank of $\lambda$ in the $(M+1)$-sample $\{\lambda^{(1)}, \dots ,\lambda^{(M)},\lambda\}$.

\section{Simulation study}\label{Simulation_study}
We simulated a cluster detection procedure in order to compare the performance of the three spatial scan statistics when data are affected by  spatial correlation: the classical scan statistic of \cite{kulldorff2009scan} that ignores the spatial correlation, and the parametric and nonparametric SAR scan statistics proposed here.
\subsection{Design of the simulation study}
Artificial datasets were generated according to the following SAR model by using the geographical locations of the $n = 94$ French administrative areas (\textit{d\'epartements}, as shown in Figure~\ref{FigCarte} in the Supplementary Material). Each location was defined as the \textit{d\'epartement}'s administrative center. Let denote by $C$ the simulated cluster defined as a set of 8 \textit{d\'epartements}. The spatial correlation is introduced by using a contiguity matrix where $w_{ii}=0$ and $w_{ij}=1$ if the two associated  \textit{d\'epartements}  are contiguous (the neighborhood graph associated to this matrix is shown in Figure~\ref{FigCarte}). The data are simulated according to the following model:      
\begin{equation}
Y_i=\rho \sum_{j=1}^{n}w_{ij} Y_j+ \alpha+\delta\times\mathbb{I}(s_i\in C) +\epsilon_i, \qquad  \epsilon_i\sim\mathcal{N}(0,\sigma^2),\qquad i=1,\ldots,n; \quad n=94.
\label{ModelSim}
\end{equation} 
Without loss of generality we took $\alpha=0$ and $\sigma=1$. The change in power of the proposed scan statistics is observed by varying both the spatial correlation intensity $\rho \in \{0, 0.2, 0.4, 0.6, 0.8\}$ and that of the simulated cluster $\delta=c\sqrt{2}$ where $c\in\{0,0.5,1,1.5\}$. Note that $\rho=0$ illustrates the situation of independent data while $c=0$ with $\rho > 0$ corresponds to the case of spatially correlated data without the presence of a cluster. 
For each pair values of $\rho$ and $c$, $S = 1000$ simulated datasets have been generated. The comparison of the three methods was performed using three distinct criteria:
the power of the method, the true-positive rate (TP) and the false-positive rate (FP). The power of each method was defined as the proportion of datasets highlighting a significant cluster, with a type I error of $0.05$ and running $999$ Monte-Carlo simulations. The TP and FP rates were calculated according to \cite{cucala2018multivariate}. It should be noted that in scenario $c=0$ the power will refer to the type I error. In this case, one will be able to observe the increase of the type I error which is only due to the spatial correlation. 

\subsection{Results}

\begin{figure}[!h] 
	\centering
	\includegraphics[width=0.6\textwidth]{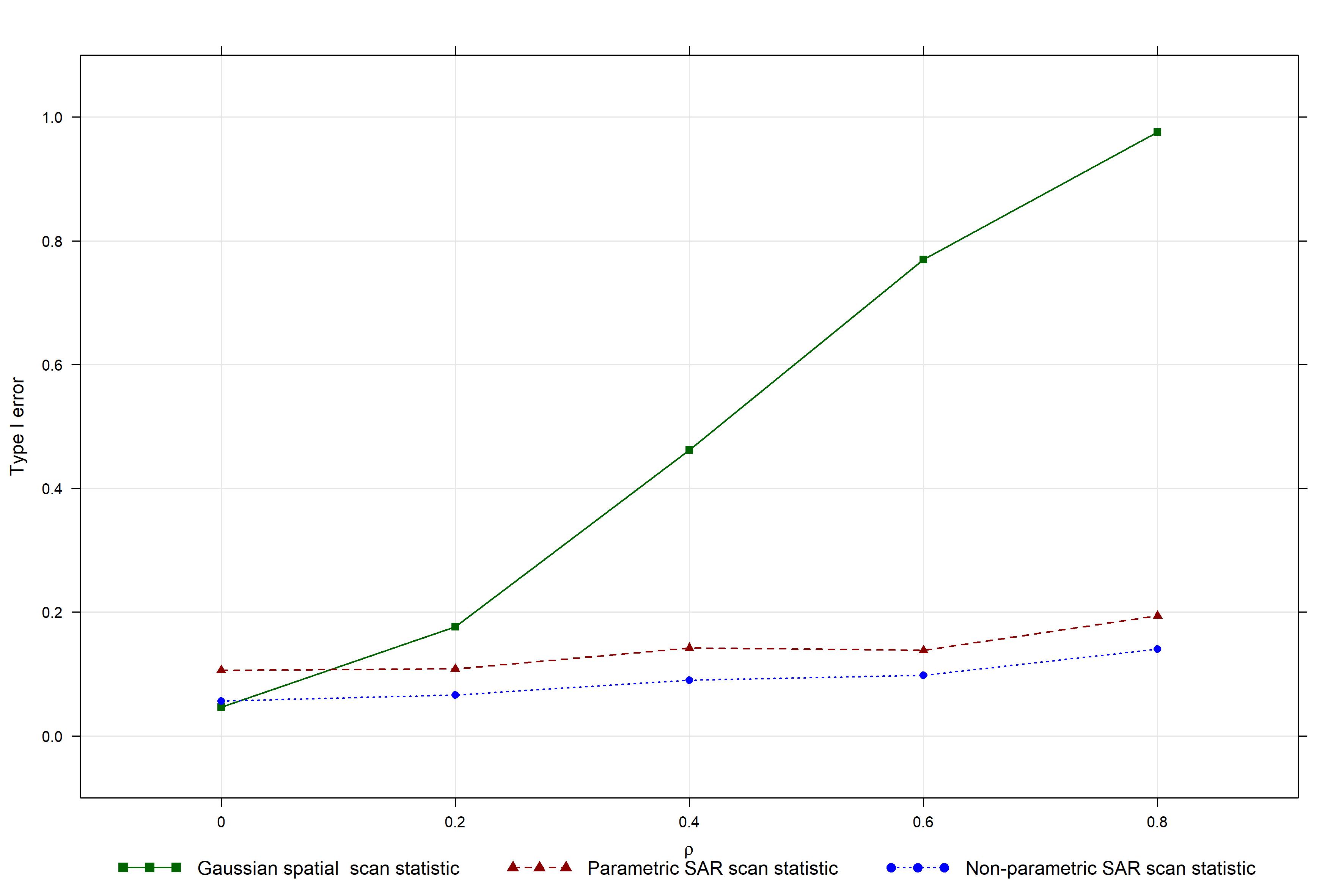}
	\caption{ Simulation study: comparison of the Gaussian spatial scan statistic and the SAR scan statistics according to type I error.  The quantity $\rho$ refers to the spatial correlation.}
	\label{Type1error}
\end{figure}

\begin{figure}[h!] 
	\centering
	\includegraphics[width=1\textwidth]{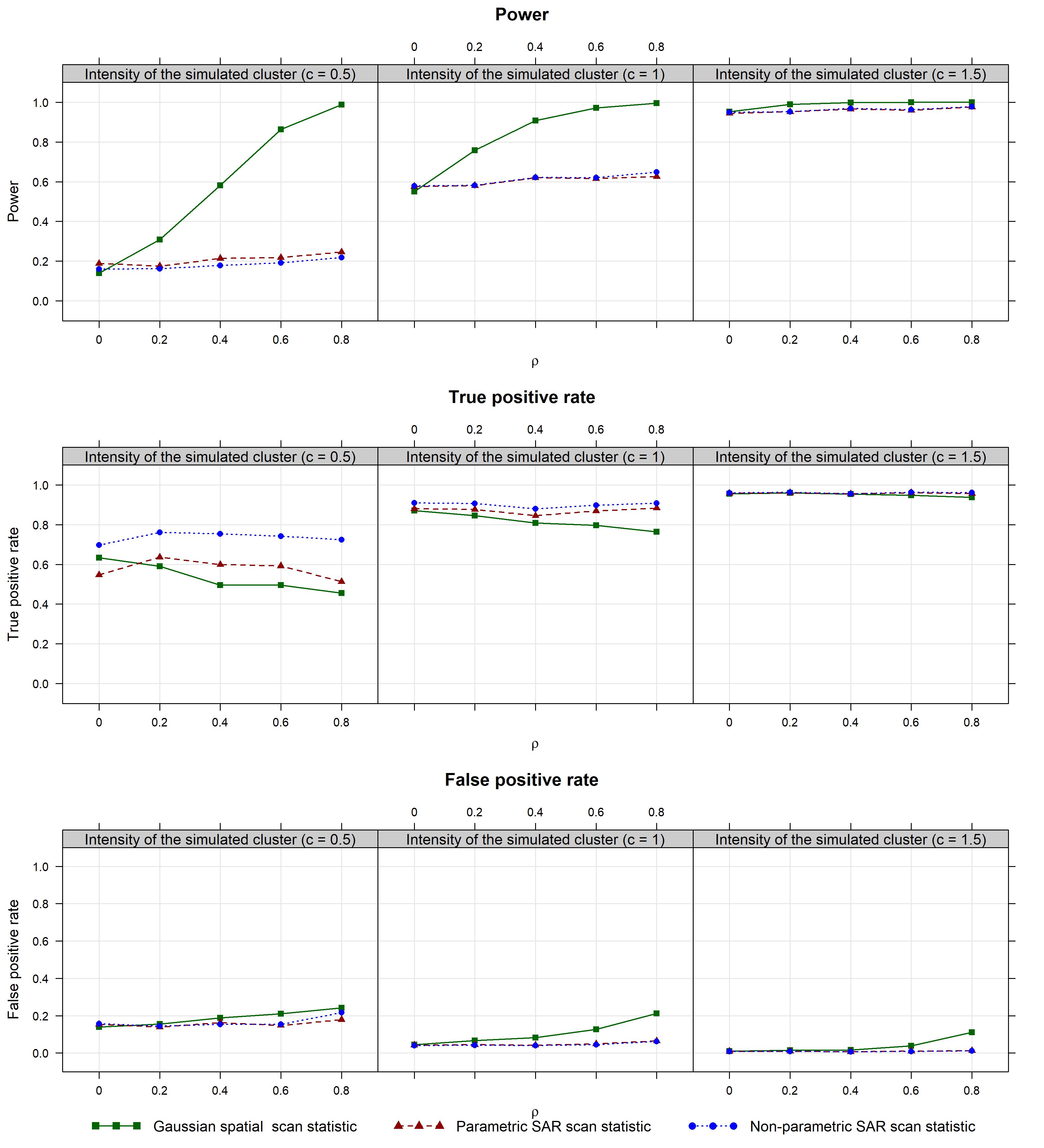}
	\caption{Simulation study: comparison of the Gaussian spatial scan statistic and the SAR scan statistics according to the cluster intensity ($c$) and spatial correlation ($\rho$). 
	For each method, the power curves and the true-positive and false-positive rates for the detection of the simulated cluster as most likely cluster are shown.}
	\label{Power}
\end{figure}

Figure \ref{Type1error} presents the comparison of the Gaussian spatial scan statistic and the parametric and nonparametric SAR scan statistics according to type I error for different values of $\rho$. Regarding the classical spatial scan statistics, the type I error sharply increases when the spatial correlation increases, while it remains approximately stable for the two others models.  
Figure \ref{Power} shows the comparison of the three methods according to power, true-positive rate and false-positive rate. 

In the absence of spatial correlation ($\rho=0$), all three models show similar powers. Both SAR scan statistics maintain this power level regardless of the intensity of the spatial correlation, while the conventional spatial scan statistic increases it with the intensity of the spatial correlation. The power increase related to the latter model is accompanied by a decrease of the true-positive rate and also an increase of the false-positive rate, (see panels associated to intensity of simulated cluster related to $c=1$). This means that the classical method has tendency to detect clusters which are shifted by spatial correlation, particularly for moderate cluster intensity and high spatial correlation ($c=1$ and $\rho=0.8$). The two SAR scan statistics show true-positive and false-positive rates that are approximately stable and close to those of the classical model associated to $\rho=0$. This demonstrates the ability of the two SAR models to obtain spatial filtered data without spatial correlation.

\subsection{Simulation study with misspecified spatial weights matrix}
\begin{figure}[!h] 
	\centering
	\includegraphics[width=1\textwidth]{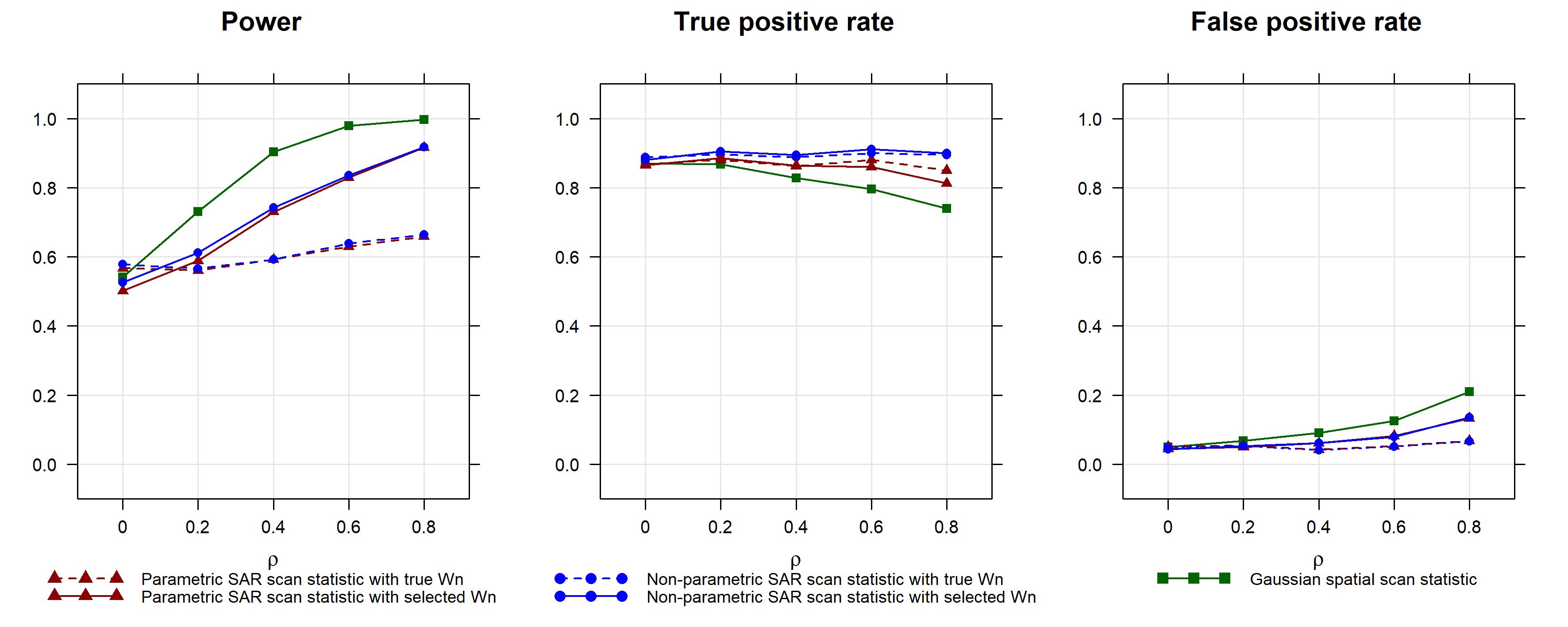}
	\caption{Simulation study: comparison of the Gaussian spatial scan statistic and the SAR scan statistics for true and selected $W_n$,  according to cluster intensity  $c=1$ and spatial correlation ($\rho$). For each method, the power curves and the true-positive and false-positive rates for the detection of the simulated cluster as most likely cluster are shown.}
	\label{misswn}
\end{figure}

The spatial weights matrix $W_n$ plays a crucial role in taking into account spatial correlation in SAR models. The specification of the elements of this weights matrix are usually considered arbitrary and the practitioner can define it by one or combination of various popular weighting schemes: $k-$NN (nearest neighbors), inverse distance, spatial contiguity,\dots (see  Chapter 3  in \cite{anselin2013spatial} for more details). In practice, we first predefine some set of candidates spatial weights matrices which may be logically suggested by the nature of the study. Then, we select the spatial weights matrix which seems to be more appropriate to the data for example the one that maximizes the Moran's I index as proposed by \cite{kooijman1976some}. It should be noted that other selection approaches are developed in the literature of spatial econometrics (see \cite{kostov2010model} for a general review).  \\
Here, we study the influence of the choice of the spatial weights matrix on the proposed SAR scan statistics. We simulate datasets from the previous SAR model (\ref{ModelSim}) where a contiguity matrix, a moderate cluster intensity ($ c = 1 $), and different spatial correlation intensities are considered ($\rho \in \{0,0.2,\dots,0.8\})$. Note that the considered contiguity matrix allows for each site to have at least two neighbors, approximately 5 neighbors on average  and a maximum of 10 neighbors. Therefore, a set of nearest  neighbors weights matrices is predefined where the number of neighbors goes from $2$ to $10$. Then, for each  simulated dataset an optimum spatial weights matrix is selected by maximizing the Moran's I index. Next, the two SAR scan statistics based on this selected spatial weights matrix are compared  with the classical spatial scan statistic. In addition, to evaluate the bias related to the misspecification of the spatial weights matrix, we compare the performance of the SAR scan statistics applied with the selected weights matrix to the the SAR scan statistics applied with the true weights matrix. \medskip\\ 
For each value of the spatial correlation intensity, $1000$ simulated datasets have been generated. The power and the TP and FP rates of each methods  have been shown  in Figure \ref{misswn}. First, it can be observed that the use of the optimal selected spatial weights matrix in SAR spatial scan statistics gives true positive rates close to those given when the true spatial weights matrix has been used. Secondly, misspecification of the spatial weights matrix leads to a small increase in the false-positive rate according to the spatial correlation intensity but which increase  remained less important to that due to the ignorance of the spatial correlation. Finally, an overpower is also observed when the spatial weights matrix is misspecified, but it is highly attenuated from the Gaussian spatial scan statistic. \\

\section{Application to economic data}\label{Real_data}
\begin{figure}[h!] 
	\centering
	\includegraphics[width=1\textwidth]{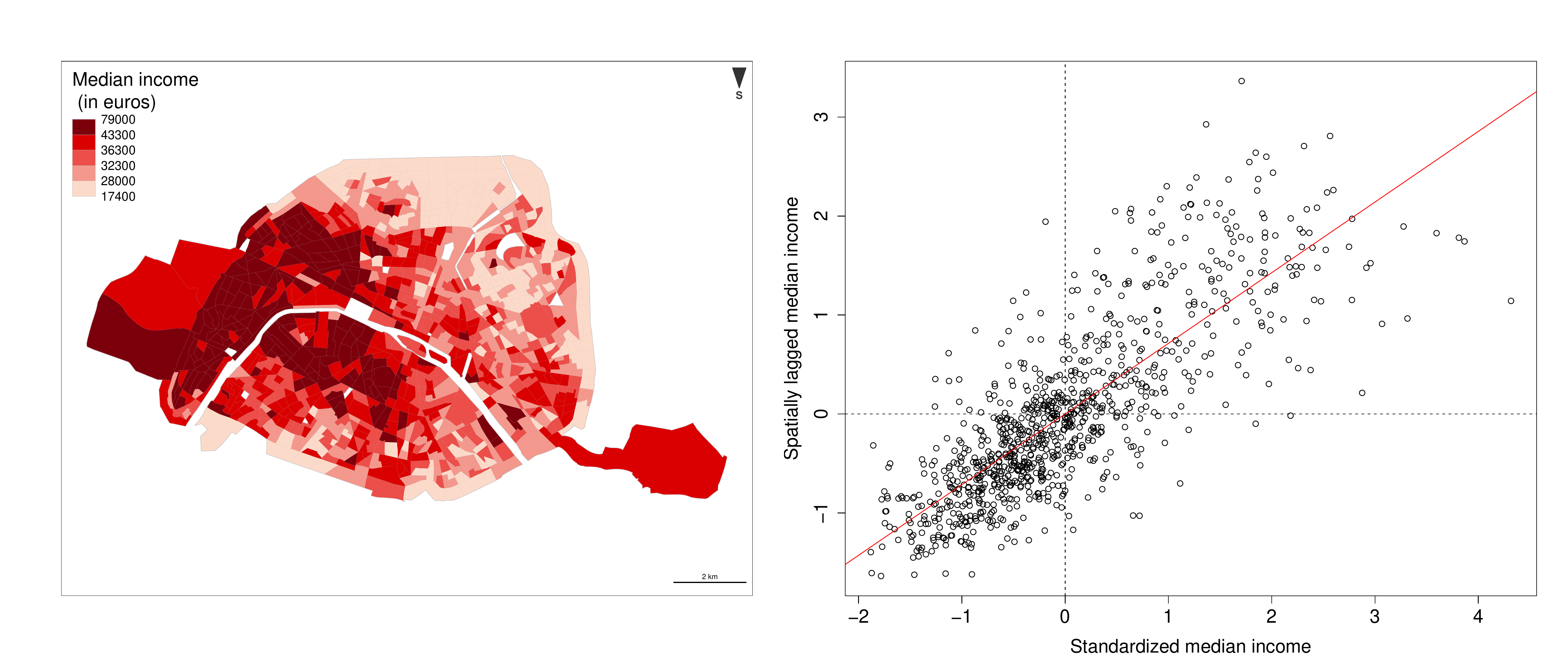}
	\caption{Median income for year 2010 in Paris and the associated Moran Scatterplot.}
	\label{revenuparis}
\end{figure}

\begin{figure}[h!] 
	\centering
	\includegraphics[width=1\textwidth]{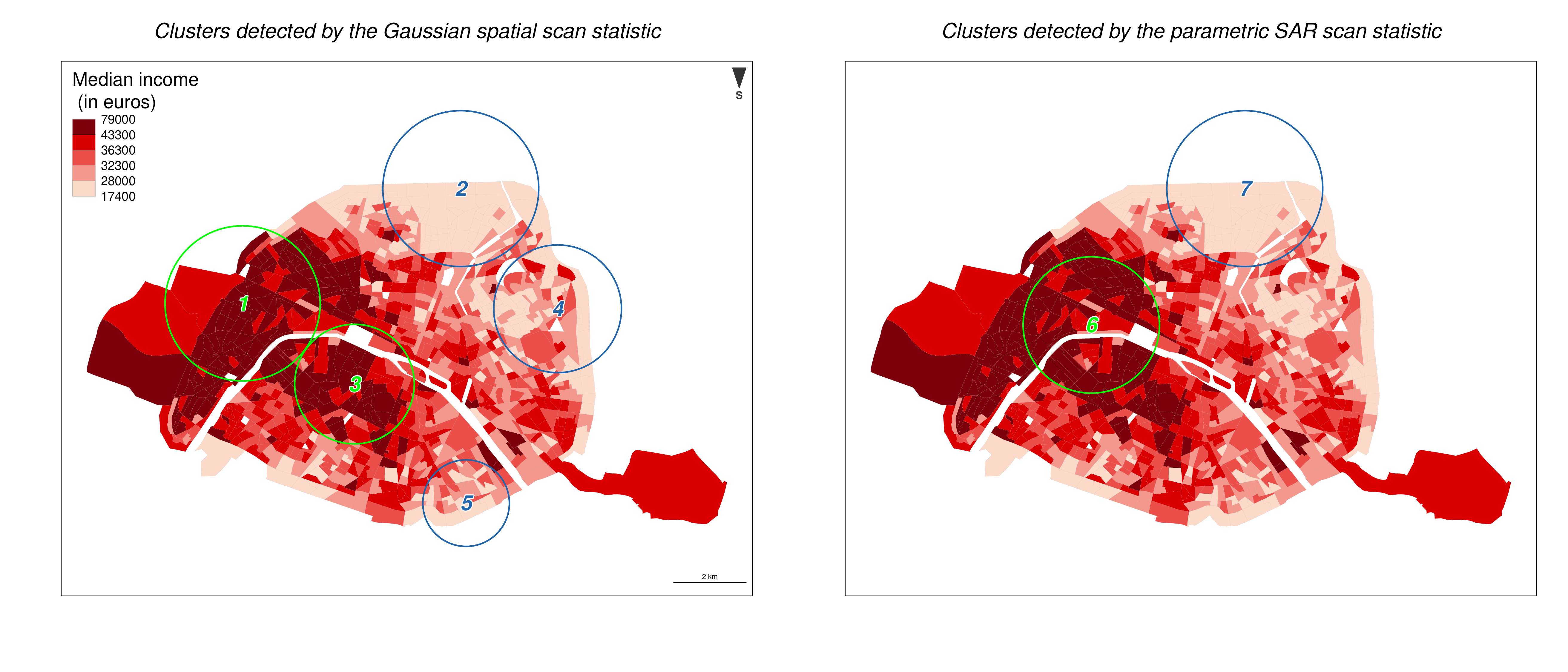}
	\caption{Statistically significant  wealthy (green circle) and unwealthy (blue circle) median income spatial clusters in Paris detected using Gaussian spatial and SAR scan statistics}
	\label{clusters}
\end{figure}

We considered data provided by the French national census database (\textit{Institut National de la Statistique et des Etudes Economiques}, INSEE) on the median income for year 2010 in each of the $953$ parisian census districts (IRIS). In Figure~\ref{revenuparis}, the spatial repartition of the median income in Paris by IRIS is illustrated in left  panel and the right panel shows the Moran's diagram. This latter is a scatter plot with the values of the standardized median income ($Y$) on the x-axis and the spatially lagged values of the standardized median income $(W_n Y)$ on the y-axis. The spatial weights matrix $W_n$ is defined by a normalized $3-$NN matrix that maximizes the Moran's I index over the all $k-$NN matrices. The Moran's diagram shows an important spatial correlation with Moran's index equal to $0.71$ and a  small p-value ($<0.001$) associated to the uncorrelated hypothesis. We define the outcome variable by the logarithm's transformation of the median income which seems to verify the normal distribution assumption, see Figure~\ref{qqplot}.\\
We aim to investigate the  existence of some clusters of wealthy and unwealthy IRIS's in Paris city and evaluate the effect of the spatial correlation on the cluster detection. Thus, statistical significant clusters detected by the Gaussian spatial scan statistic are compared to those detected by the SAR scan statistics. The SAR scan statistics have been used with the selected $3$ nearest neighbors matrix. The statistical significance of detected clusters was evaluated by performing $999$ Monte-Carlo simulations, considering a type I error of $0.05$. Remark that we are able to detect multiple non-overlapping clusters by using the sequential detection approach proposed by \cite{leespatial}. \\
The results of the Gaussian spatial and  the SAR scan statistics are presented in Table~\ref{tabIris}  and Figure~\ref{clusters}. It should be noted that the parametric and non-parametric SAR scan statistics have given the same results, which is not surprising since the distribution of the outcome variable looks quite Gaussian. For this reason only the parametric SAR scan statistic is illustrated here.\\
The Gaussian spatial scan statistic identified $5$ significant clusters: 2 wealthy clusters (cluster 1 and 3) and 3 unwealthy clusters (cluster 2, 4, and 5), while the SAR scan statistic identified only two significant clusters: a wealthy cluster (cluster 6) and an unwealthy cluster (cluster 7).\\
First, remark that the unwealthy clusters 2 and 7 are exactly the same: both methods identify this area, next to the French \textit{d\'epartement} with the highest poverty rate (Seine-Saint-Denis), as the most unwealthy in Paris. Then we may notice that two unwealthy clusters detected by the Gaussian spatial scan statistic are no longer significant once the effect of spatial correlation has been removed. Finally, the wealthy clusters 1 and 3 detected by the Gaussian spatial scan statistic have been shifted to the thinner cluster 6 when using the SAR scan statistic. This phenomenon is similar to what we observed in the simulation study. It should be noted that the wealthy cluster 6 is centered almost exactly on the location of Eiffel Tower, symbol of Paris. Unsurprisingly, wealthy people tend to aggregate around this monument and it becomes more obvious when taking into account the spatial correlation.

\begin{table}
\centering
\label{tabIris}
\caption{Statistically significant spatial clusters of median income in Paris detected by the Gaussian spatial and SAR scan statistics.}
{\footnotesize
	\begin{tabular}{c c c c c ccc}
\hline
Model & Cluster & \# IRIS & Mean inside & SD inside & Mean outside & SD outside& p-value\\\hline 

Gaussian &	1	&	94	&	50565	&	9259	&	34394	&	8555	&	0.001\\
scan&	2	&	88	&	26220	&	4224	&	36983	&	9749	&	0.001\\
statistic&	3	&	77	&	45886	&	8177	&	35119	&	9540	&	0.001 \\
&	4	&	95	&	28323	&	4909	&	36838	&	9927	&	0.001 \\
&	5	&	36	&	28541	&	4550	&	36282	&	9919	&	0.015\\ \hline
SAR scan&	6	&	95	&	49379	&	9326	&	34507	&	8764	&	0.001\\
statistic&	7	&	88	&	26220	&	4225	&	36983	&	9749	&	0.003\\
  \hline 
	\end{tabular}
	}
\end{table}

\section{Discussion}\label{Discussion}
In this paper, we developed parametric and nonparametric SAR scan statistics for continuous spatial data in order to take into consideration the spatial correlation which is generally present in spatial data. In a first time, we proposed to account for the spatial correlation of the outcome variable by using an easy-to-implement spatial filtration method based on SAR models. In a second time, we proposed to apply a parametric or non-parametric spatial scan statistic (depending on the distribution of initial data) on the spatially filtrated outcome. \\
Through a simulation study, the performance of the proposed SAR scan statistics and that of the classical spatial scan statistic were compared. The simulation results showed that the ignorance of the spatial correlation increases the type I error and false positive rate. In addition, it decreases the true-positive rate by detecting spatial clusters shifted by spatial correlation. In contrast, the SAR scan statistics allowed to keep the level of the type I error, the true and false positive rates regardless of the intensity of the spatial correlation, particularly for the case of well-specified spatial weights matrix. \\
The choice of the spatial weights matrix was evaluated through simulation study.  This latter highlighted that the effects related to the misspecification of this matrix on the proposed approaches has to be considered. Hence, we proposed a simple method to choose the optimal weights matrix using Moran's index. It should be noted that others more sophisticated methods for selecting the spatial weights matrix, likes that of \cite{kostov2010model}, can be used in the proposed SAR scan statistics. \\
The SAR scan statistics were applied  to the median income for year 2010 at each of the $953$ parisian census districts (IRIS) in order to detect wealthy and unwealthy clusters not affected by the spatial correlation. Compared to the Gaussian spatial scan statistic, the SAR scan statistics have detected more concentrated clusters. 
Cluster detection is usually adjusted for confounding factors. The proposed SAR scan statistics rely on the SAR models and the scan statistic approach of \cite{jung2009generalized} which was initially developed for adjusting for confounding  factors. However, confounding  factors can be easily added in model (\ref{modele1})  and theirs coefficients have to be estimated with the spatial autoregressive parameter. Then theirs effects have to be integrated as an offset in the spatially filtered outcome.\\
Recently, \cite{cucala2017multivariate} designed a Gaussian scan statistic for spatial multivariate data. Their method takes into account the correlation between different variables but assumes independence between neighbouring sites. Adapting this method to account for spatial correlation could be highly interesting, but also quite challenging since the spatial autocorrelation parameter might be different from one variable to another.



%
%
%
%

\bibliographystyle{model2-names}
\bibliography{biblio}

\begin{thebibliography}{24}
\expandafter\ifx\csname natexlab\endcsname\relax\def\natexlab#1{#1}\fi
\providecommand{\url}[1]{\texttt{#1}}
\providecommand{\href}[2]{#2}
\providecommand{\path}[1]{#1}
\providecommand{\DOIprefix}{doi:}
\providecommand{\ArXivprefix}{arXiv:}
\providecommand{\URLprefix}{URL: }
\providecommand{\Pubmedprefix}{pmid:}
\providecommand{\doi}[1]{\href{http://dx.doi.org/#1}{\path{#1}}}
\providecommand{\Pubmed}[1]{\href{pmid:#1}{\path{#1}}}
\providecommand{\bibinfo}[2]{#2}
\ifx\xfnm\relax \def\xfnm[#1]{\unskip,\space#1}\fi
\bibitem[{Anselin(2013)}]{anselin2013spatial}
\bibinfo{author}{Anselin, L.}, \bibinfo{year}{2013}.
\newblock \bibinfo{title}{Spatial econometrics: methods and models}.
  volume~\bibinfo{volume}{4}.
\newblock \bibinfo{publisher}{Springer Science \& Business Media}.
\bibitem[{Bhatt and Tiwari(2014)}]{bhatt2014spatial}
\bibinfo{author}{Bhatt, V.}, \bibinfo{author}{Tiwari, N.},
  \bibinfo{year}{2014}.
\newblock \bibinfo{title}{A spatial scan statistic for survival data based on
  weibull distribution}.
\newblock \bibinfo{journal}{Statistics in medicine} \bibinfo{volume}{33},
  \bibinfo{pages}{1867--1876}.
\bibitem[{Burnham and Anderson(2004)}]{burnham2004multimodel}
\bibinfo{author}{Burnham, K.P.}, \bibinfo{author}{Anderson, D.R.},
  \bibinfo{year}{2004}.
\newblock \bibinfo{title}{Multimodel inference: understanding aic and bic in
  model selection}.
\newblock \bibinfo{journal}{Sociological methods \& research}
  \bibinfo{volume}{33}, \bibinfo{pages}{261--304}.
\bibitem[{Cliff and Ord(1973)}]{Cliff}
\bibinfo{author}{Cliff, A.}, \bibinfo{author}{Ord, K.}, \bibinfo{year}{1973}.
\newblock \bibinfo{title}{Spatial autocorrelation}.
\newblock \bibinfo{journal}{London: Pion Ltd} .
\bibitem[{Cressie(1977)}]{Cressie1977}
\bibinfo{author}{Cressie, N.}, \bibinfo{year}{1977}.
\newblock \bibinfo{title}{On some properties of the scan statistic on the
  circle and the line}.
\newblock \bibinfo{journal}{Journal of Applied Probability}
  \bibinfo{volume}{14}, \bibinfo{pages}{272--283}.
\bibitem[{Cucala(2014)}]{cucala2014distribution}
\bibinfo{author}{Cucala, L.}, \bibinfo{year}{2014}.
\newblock \bibinfo{title}{A distribution-free spatial scan statistic for marked
  point processes}.
\newblock \bibinfo{journal}{Spatial Statistics} \bibinfo{volume}{10},
  \bibinfo{pages}{117--125}.
\bibitem[{Cucala et~al.(2013)Cucala, Demattei, Lopes and
  Ribeiro}]{cucala2013spatial}
\bibinfo{author}{Cucala, L.}, \bibinfo{author}{Demattei, C.},
  \bibinfo{author}{Lopes, P.}, \bibinfo{author}{Ribeiro, A.},
  \bibinfo{year}{2013}.
\newblock \bibinfo{title}{A spatial scan statistic for case event data based on
  connected components}.
\newblock \bibinfo{journal}{Computational Statistics} \bibinfo{volume}{28},
  \bibinfo{pages}{357--369}.
\bibitem[{Cucala et~al.(2017)Cucala, Genin, Lanier and
  Occelli}]{cucala2017multivariate}
\bibinfo{author}{Cucala, L.}, \bibinfo{author}{Genin, M.},
  \bibinfo{author}{Lanier, C.}, \bibinfo{author}{Occelli, F.},
  \bibinfo{year}{2017}.
\newblock \bibinfo{title}{A multivariate gaussian scan statistic for spatial
  data}.
\newblock \bibinfo{journal}{Spatial Statistics} \bibinfo{volume}{21},
  \bibinfo{pages}{66--74}.
\bibitem[{Cucala et~al.(2019)Cucala, Genin, Occelli and
  Soula}]{cucala2018multivariate}
\bibinfo{author}{Cucala, L.}, \bibinfo{author}{Genin, M.},
  \bibinfo{author}{Occelli, F.}, \bibinfo{author}{Soula, J.},
  \bibinfo{year}{2019}.
\newblock \bibinfo{title}{A multivariate nonparametric scan statistic for
  spatial data}.
\newblock \bibinfo{journal}{Spatial Statistics} \bibinfo{volume}{29},
  \bibinfo{pages}{1--14}.
\bibitem[{Huang et~al.(2007)Huang, Kulldorff and Gregorio}]{huang2007spatial}
\bibinfo{author}{Huang, L.}, \bibinfo{author}{Kulldorff, M.},
  \bibinfo{author}{Gregorio, D.}, \bibinfo{year}{2007}.
\newblock \bibinfo{title}{A spatial scan statistic for survival data}.
\newblock \bibinfo{journal}{Biometrics} \bibinfo{volume}{63},
  \bibinfo{pages}{109--118}.
\bibitem[{Huang et~al.(2009)Huang, Tiwari, Zou, Kulldorff and
  Feuer}]{huang2009weighted}
\bibinfo{author}{Huang, L.}, \bibinfo{author}{Tiwari, R.C.},
  \bibinfo{author}{Zou, Z.}, \bibinfo{author}{Kulldorff, M.},
  \bibinfo{author}{Feuer, E.J.}, \bibinfo{year}{2009}.
\newblock \bibinfo{title}{Weighted normal spatial scan statistic for
  heterogeneous population data}.
\newblock \bibinfo{journal}{Journal of the American Statistical Association}
  \bibinfo{volume}{104}, \bibinfo{pages}{886--898}.
\bibitem[{Jung(2009)}]{jung2009generalized}
\bibinfo{author}{Jung, I.}, \bibinfo{year}{2009}.
\newblock \bibinfo{title}{A generalized linear models approach to spatial scan
  statistics for covariate adjustment}.
\newblock \bibinfo{journal}{Statistics in medicine} \bibinfo{volume}{28},
  \bibinfo{pages}{1131--1143}.
\bibitem[{Jung and Cho(2015)}]{jung2015nonparametric}
\bibinfo{author}{Jung, I.}, \bibinfo{author}{Cho, H.J.}, \bibinfo{year}{2015}.
\newblock \bibinfo{title}{A nonparametric spatial scan statistic for continuous
  data}.
\newblock \bibinfo{journal}{International journal of health geographics}
  \bibinfo{volume}{14}, \bibinfo{pages}{30}.
\bibitem[{Kooijman(1976)}]{kooijman1976some}
\bibinfo{author}{Kooijman, S.}, \bibinfo{year}{1976}.
\newblock \bibinfo{title}{Some remarks on the statistical analysis of grids
  especially with respect to ecology}, in: \bibinfo{booktitle}{Annals of
  Systems Research}. \bibinfo{publisher}{Springer}, pp.
  \bibinfo{pages}{113--132}.
\bibitem[{Kostov(2010)}]{kostov2010model}
\bibinfo{author}{Kostov, P.}, \bibinfo{year}{2010}.
\newblock \bibinfo{title}{Model boosting for spatial weighting matrix selection
  in spatial lag models}.
\newblock \bibinfo{journal}{Environment and Planning B: Planning and Design}
  \bibinfo{volume}{37}, \bibinfo{pages}{533--549}.
\bibitem[{Kulldorff(1997)}]{kulldorff1997spatial}
\bibinfo{author}{Kulldorff, M.}, \bibinfo{year}{1997}.
\newblock \bibinfo{title}{A spatial scan statistic}.
\newblock \bibinfo{journal}{Communications in Statistics-Theory and methods}
  \bibinfo{volume}{26}, \bibinfo{pages}{1481--1496}.
\bibitem[{Kulldorff et~al.(2009)Kulldorff, Huang and Konty}]{kulldorff2009scan}
\bibinfo{author}{Kulldorff, M.}, \bibinfo{author}{Huang, L.},
  \bibinfo{author}{Konty, K.}, \bibinfo{year}{2009}.
\newblock \bibinfo{title}{A scan statistic for continuous data based on the
  normal probability model}.
\newblock \bibinfo{journal}{International journal of health geographics}
  \bibinfo{volume}{8}, \bibinfo{pages}{58}.
\bibitem[{Kulldorff et~al.(2006)Kulldorff, Huang, Pickle and
  Duczmal}]{kulldorff2006elliptic}
\bibinfo{author}{Kulldorff, M.}, \bibinfo{author}{Huang, L.},
  \bibinfo{author}{Pickle, L.}, \bibinfo{author}{Duczmal, L.},
  \bibinfo{year}{2006}.
\newblock \bibinfo{title}{An elliptic spatial scan statistic}.
\newblock \bibinfo{journal}{Statistics in medicine} \bibinfo{volume}{25},
  \bibinfo{pages}{3929--3943}.
\bibitem[{Lee et~al.(2019)Lee, Sun and Chang}]{leespatial}
\bibinfo{author}{Lee, J.}, \bibinfo{author}{Sun, Y.}, \bibinfo{author}{Chang,
  H.H.}, \bibinfo{year}{2019}.
\newblock \bibinfo{title}{Spatial cluster detection of regression coefficients
  in a mixed-effects model}.
\newblock \bibinfo{journal}{Environmetrics} , \bibinfo{pages}{e2578}.
\bibitem[{Lee(2004)}]{lee2004asymptotic}
\bibinfo{author}{Lee, L.F.}, \bibinfo{year}{2004}.
\newblock \bibinfo{title}{Asymptotic distributions of quasi-maximum likelihood
  estimators for spatial autoregressive models}.
\newblock \bibinfo{journal}{Econometrica} \bibinfo{volume}{72},
  \bibinfo{pages}{1899--1925}.
\bibitem[{LeSage and Pace(2009)}]{lesage2009introduction}
\bibinfo{author}{LeSage, J.}, \bibinfo{author}{Pace, R.K.},
  \bibinfo{year}{2009}.
\newblock \bibinfo{title}{Introduction to spatial econometrics}.
\newblock \bibinfo{publisher}{Chapman and Hall/CRC}.
\bibitem[{Lin(2014)}]{lin2014generalized}
\bibinfo{author}{Lin, P.S.}, \bibinfo{year}{2014}.
\newblock \bibinfo{title}{Generalized scan statistics for disease
  surveillance}.
\newblock \bibinfo{journal}{Scandinavian Journal of Statistics}
  \bibinfo{volume}{41}, \bibinfo{pages}{791--808}.
\bibitem[{Loh and Zhu(2007)}]{loh2007accounting}
\bibinfo{author}{Loh, J.M.}, \bibinfo{author}{Zhu, Z.}, \bibinfo{year}{2007}.
\newblock \bibinfo{title}{Accounting for spatial correlation in the scan
  statistic}.
\newblock \bibinfo{journal}{The Annals of Applied Statistics}
  \bibinfo{volume}{1}, \bibinfo{pages}{560--584}.
\bibitem[{Raftery(1995)}]{raftery1995bayesian}
\bibinfo{author}{Raftery, A.E.}, \bibinfo{year}{1995}.
\newblock \bibinfo{title}{Bayesian model selection in social research}.
\newblock \bibinfo{journal}{Sociological methodology} \bibinfo{volume}{25},
  \bibinfo{pages}{111--164}.

\end{thebibliography}
\appendix
\label{Appendix}
\section{Explicit expressions of the parameters estimators for the Gaussian spatial scan statistics}
Under $\mathcal{H}_0$, the MLEs of $\alpha$ and $\sigma^2$ have the following explicit expressions:
 \begin{equation*}
 \widehat{\alpha}=\frac{1}{n}\sum_{i=1}^{n}Y_i\qquad \mbox{and} \qquad \widehat{\sigma^2}=\frac{1}{n}\sum_{i=1}^{n}\left( Y_i-\widehat{\alpha}\right)^{2}. 
 \end{equation*}
Under $\mathcal{H}_1$, the MLEs of $\alpha$, $\sigma^2$ and $\delta_k$ have the following explicit expressions:

\begin{equation*}
\widehat{\alpha}_{k}=\frac{1}{n-n_k}\sum_{i=1}^{n}\left(1-\xi^{(k)}_i\right)Y_i,\qquad \widehat{\delta}_{k}=\frac{1}{n-n_k}\sum_{i=1}^{n}\left(\frac{n}{n_k}\xi^{(k)}_i-1\right){Y}_i
\end{equation*}
and 
\begin{eqnarray*}
	\widehat{\sigma^2}_{k}&=&\frac{1}{n}\sum_{i=1}^{n}\left(Y_i-\widehat{\alpha}_{k}-\widehat{\delta}_{k}\xi_{i}^{(k)}\right)^2\\ 
	&=&  \frac{1}{n}\left\{\sum_{i \notin C_k}\left(Y_i-\frac{1}{n-n_k}\sum_{j \notin C_k }Y_j\right)^2 +\sum_{i\in C_k}\left(Y_i-\frac{1}{n_k}\sum_{j \in C_k}Y_j\right)^2\right\}
\end{eqnarray*}
where $n_k$ is the number of locations inside $C_k$. Thus, the last decomposition is equal to the estimator of $\sigma^2$ under $\mathcal{H}_1$ given in \cite{kulldorff2009scan}.

\begin{figure}[!h]
\centering
\includegraphics[width=0.5\textwidth]{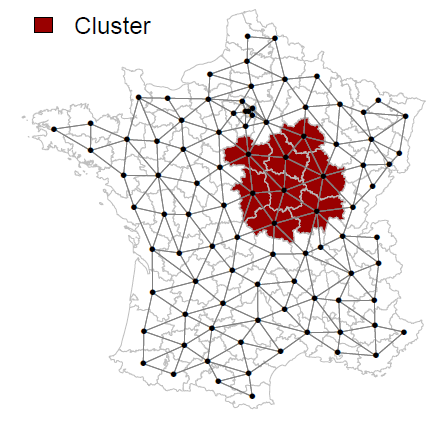}

\caption{Neighborhood graph and simulated cluster. }
\label{FigCarte}
\end{figure}

\begin{figure}[!h]
	\centering
	\includegraphics[width=0.5\textwidth]{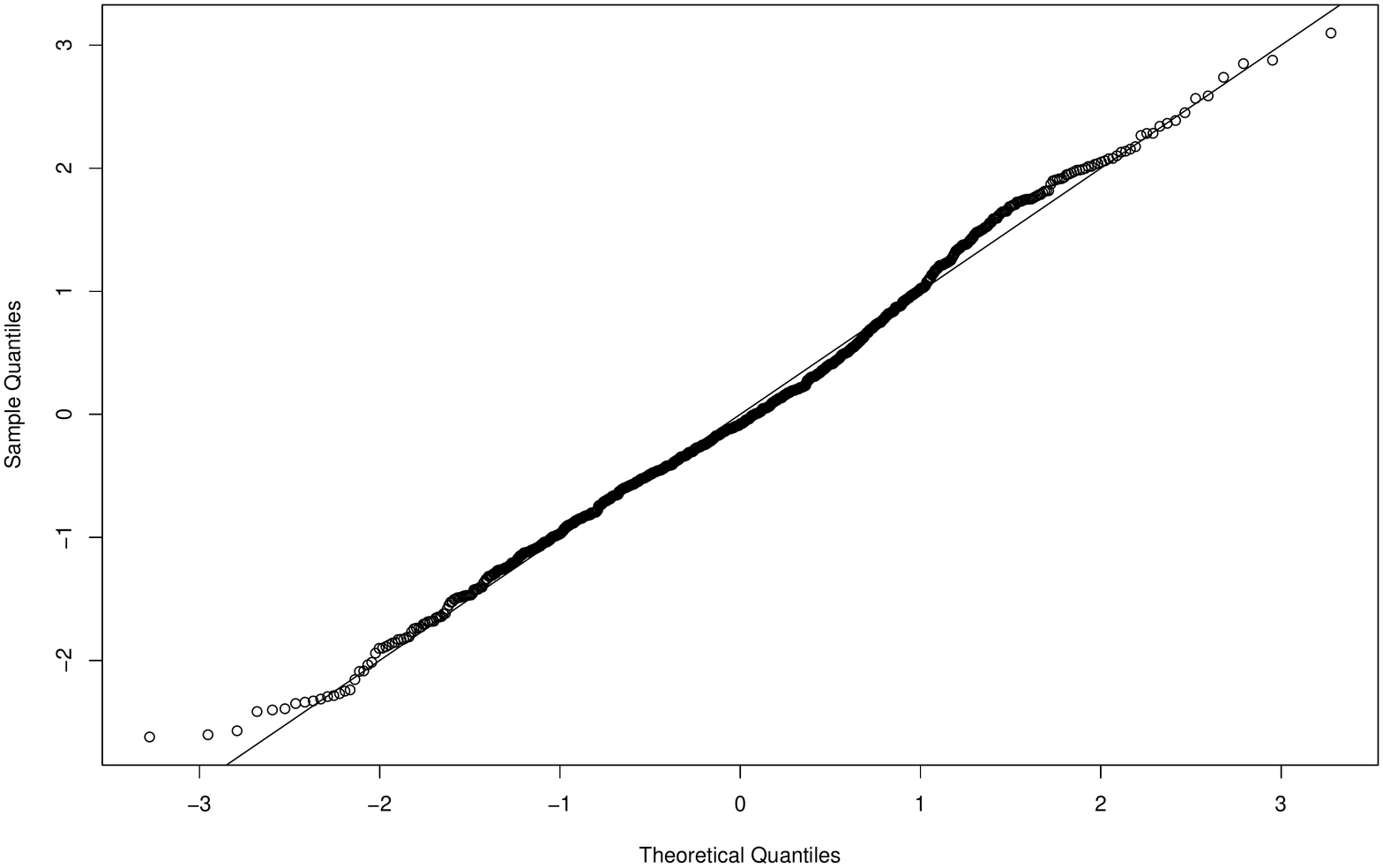}
	
	\caption{Normal Q-Q plot of the standardized median income of 2010 in  Paris. }
	\label{qqplot}
\end{figure}


%
%

\end{document}